\documentclass[aps,prl,floatfix,twocolumn,showpacs,10pt,longbibliography]{revtex4-2}
\usepackage{amsmath}
\usepackage{url}
\usepackage[ruled,vlined]{algorithm2e}
\usepackage{algorithmic}
\usepackage{graphicx}
\usepackage{dcolumn}
\usepackage{bm}
\usepackage{amssymb}
\usepackage{rotating}
\usepackage[abs]{overpic}
\usepackage{xcolor}
\usepackage{hyperref}
\usepackage{verbatim}
\hypersetup{colorlinks = true, citebordercolor={blue}, linkcolor={blue}, citecolor={blue}, urlcolor={blue}}

\newcommand{\tc}{\textcolor{black}}

\begin{document}




\title{Large Cumulant Eigenvalue as a Signature of Exciton Condensation}

\author{Anna O. Schouten, LeeAnn M. Sager-Smith, and David A. Mazziotti}

\email{damazz@uchicago.edu}

\affiliation{Department of Chemistry and The James Franck Institute, The University of Chicago, Chicago, IL 60637 USA}%

\date{Submitted November 5, 2021\tc{; Revised January 17, 2022}\textcolor{black}{; Revised May 16, 2022}}


\pacs{31.10.+z}

\begin{abstract}
The Bose-Einstein condensation of excitons into a single quantum state is known as exciton condensation.  Exciton condensation, which potentially supports the frictionless flow of energy, has recently been realized in graphene bilayers and van der Waals heterostructures.  Here we show that exciton condensates can be predicted from a combination of reduced density matrix theory and cumulant theory. We show that exciton condensation occurs if and only if there exists a large eigenvalue in the cumulant part of the particle-hole reduced density matrix. In the thermodynamic limit we show that the large eigenvalue is bounded from above by the number of excitons.  In contrast to the eigenvalues of the particle-hole matrix, the large eigenvalue of the cumulant matrix has the advantage of providing a size-extensive measure of the extent of condensation.  Here we apply this signature to predict exciton condensation in both the \color{black} Lipkin \color{black} model and molecular stacks of benzene.  The computational signature has applications to the prediction of exciton condensation in both molecules and materials.

\end{abstract}

\maketitle

\section{Introduction}
Exciton condensation---condensation of particle-hole pairs to a single quantum state---has captured considerable experimental and theoretical interest because of the potential applications to energy efficiency. The phenomenon occurs through a type of Bose-Einstein condensation (BEC) and exhibits superfluidity \cite{london_1938,tisza_1947,DMA1995,AEM1995}. Rather than transferring charge or mass, superfluid transfer of excitons allows for dissipationless transport of energy \cite{Keldysh2017, Fil2018}, which presents the possibility for uniquely energy efficient materials. Unfortunately, exciton condensation has proven difficult to realize experimentally; however, evidence has been found in polaritons (excitons coupled to photons) \cite{KRK2006,fuhrer_hamilton_2016, PannirSivajothi2021}, and quantum wells formed from two-dimensional structures like semiconductors\cite{Butov1994, Ma2021}, graphene bilayers \cite{Liu2017, Min2008}, and van der Waals heterostructures \cite{Sigl2020,Wang2019,Fogler2014, Kogar2017}. Theoretical exploration using a signature found in the particle-hole reduced density matrix (RDM) has also revealed that exciton condensation is possible in quantum and molecular systems \cite{Safaei2018,Sager2020,sager_53, Sager_coronene, Schouten2021a}.

Information about many molecular properties and characteristics related to correlation is contained in the two-electron reduced density matrix (2-RDM) and related particle-hole reduced density matrix (RDM). \textcolor{black}{Density matrices are formed by multiplication of an $N$-electron wavefunction $| \Psi \rangle$ by the adjoint $\langle \Psi |$, yielding a matrix that contains the same molecular information as the original wavefunction \cite{M2007}. The 2-RDM, which comes from integration of the density matrix over all but two electrons,  can be expanded to be written as a wedge-product of the one electron RDMs (1-RDMs) and a cumulant, or ``connected", part \cite{Mazziotti1998a, Mazziotti1998}}. The cumulant, which cannot be written as a product of the lower-order RDMs, \tc{is size extensive in that it scales linearly with system size} \cite{Mazziotti1998a, Mazziotti2012}. Unlike its cumulant part, the 2-RDM is not size extensive because it contains \tc{the wedge product of the 1-RDM with itself which scales quadratically}.  Cumulants are considered to be the ``connected” part of the RDM because they disappear in the absence of correlation between the 1-electron matrix variables \cite{Bianucci2020, Mazziotti1998a}. \textcolor{black}{A wavefunction is correlated (meaning, that it cannot be written as a single Slater determinant) if and only if its 2-RDM's cumulant is non-vanishing.  Correlation corresponds to entanglement, or off-diagonal order, meaning the $N$-electron density matrix cannot be written as a product of 1-electron density matrices.} Much of the correlation information contained in the RDM can be obtained without the complete 2-RDM from the cumulant. In quantum chemistry and condensed-matter physics, cumulants have important applications because of \tc{their size-extensive properties} and their inherent connection to correlation. Cumulants have been used to measure many types of correlation and entanglement in molecular systems \cite{Alcoba2010, Harriman2007, Juhasz2006, Skolnik2013, Pavlyukh2014, Werba2019, RamosCordoba2016}, including those associated with superconductivity \cite{Raeber2015, Schmidt2019}. Fermion-pair condensation, a requirement for Bardeen-Copper-Schrieffer superconductivity, is detectable in the 2-RDM with a definite signature, the appearance of a large eigenvalue \cite{Yang1962, Sasaki1965}.  \textcolor{black}{The large eigenvalue, arising from off-diagonal long-range order in the 2-RDM in which fermion pairs are strongly coupled, corresponds to the population of fermion pairs in the condensate.} Raeber and Mazziotti \cite{Raeber2015} showed that the same signature of condensation also exists in the cumulant of the 2-RDM. Moreover, the eigenvalues of the cumulant and the 2-RDM share the same upper bound in the thermodynamic limit.

The signature of fermion-pair condensation in the 2-RDM has an analogous signature of exciton condensation in the particle-hole RDM. In the particle-hole RDM, \tc{one large eigenvalue trivially corresponds to the action of the number operator applied to the ground state while the appearance of} an \tc{additional} large eigenvalue reveals \tc{collective excitations that are also known as exciton condensation, with} the magnitude indicating the extent of condensation \cite{Garrod1969, Safaei2018}. Here, we explore the large eigenvalue of the particle-hole cumulant as a measure of exciton condensation. We derive a relationship between the \tc{non-trivial large eigenvalue} of the particle-hole RDM and the large eigenvalue of the particle-hole cumulant, showing that the cumulant captures the signature of exciton condensation. Additionally, we use this relationship to find an upper bound on the value of the large eigenvalue of the cumulant. \tc{Importantly, the large eigenvalue of the cumulant RDM has a significant advantage over the non-trivial large eigenvalue of the particle-hole RDM because unlike the particle-hole RDM, the cumulant RDM is size extensive. For example, for two exciton condensates separated by an infinite distance, the large eigenvalue of the cumulant RDM is the sum of the large eigenvalues of each of the condensates while the equivalent eigenvalue of the non-size-extensive particle-hole RDM is generally unequal to the sum.}  Finally, we examine this relationship in quantum and molecular systems using the Lipkin model and a system of stacked benzene molecules. We use the Lipkin model\color{black}---a standard test-Hamiltonian first proposed in the context of nuclear physics \cite{Lipkin_model,Stein_2000,Co_2018} that is capable of demonstrating the maximal degree of exciton condensation \cite{sager_coex,sager_53}---\color{black} to prepare an exciton condensate on \color{black} IBM's QASM Simulator, which conducts probabilistic classical computations that approximate an ideal quantum device, \color{black} and demonstrate that the results are consistent with the theory we derive. Using variational 2-RDM theory \cite{Mazziotti1998, Mazziotti2012, Mazziotti2004, mazziotti_2016, nakata_nakatsuji_ehara_fukuda_nakata_fujisawa_2001, M2007}, we explore the eigenvalues of the particle-hole RDM and the particle-hole cumulant for stacked benzene molecules and find that the correlation present in the particle-hole RDM is also present in the particle-hole cumulant.

\section{Theory}

We begin by relating the cumulant to the particle-hole RDM for a quantum system of $N$ fermionic particles in a finite basis set of $r$ orbitals. The elements of the particle-hole RDM
\color{black}
\begin{equation}
    ^{2} G^{i,j}_{k,l} = \langle \Psi | {\hat a}^{\dagger}_{i} {\hat a}_{j} {\hat a}^{\dagger}_{l} {\hat a}_{k} | \Psi \rangle
\end{equation}
\color{black}
can be written in terms of the elements of the 2-RDM
 \color{black}
\begin{equation}
    ^{2} D^{i,l}_{k,j} = \langle \Psi | {\hat a}^{\dagger}_{i} {\hat a}^{\dagger}_{l} {\hat a}_{j} {\hat a}_{k} | \Psi \rangle
\end{equation}
where ${\hat a}^{\dagger}_{i}$ and ${\hat a}_{i}$ are the second-quantized creation and annihilation operators for a fermion in orbital $i$ and $| \Psi \rangle$ is the wavefunction.
\color{black}
Using a simple transformation from the rearrangement of second-quantized operators~\cite{Mazziotti2012} yields
 \color{black}
\begin{equation}
\label{eq:g}
    {}^{2}G^{i,j}_{k,l} = ^{1} I^{j}_{l} \, {}^{1} D^{i}_{k} - {}^{2} D^{i,l}_{k,j} .
\end{equation}
\color{black}
where $^{1}D$ is the 1-RDM, $^{1}I$ is the identity matrix, and the integer indices $i,j,k,l \in [1,r]$ correspond to the orbitals.
Combining this transformation with the cumulant expansion of the 2-RDM~\cite{Mazziotti1998a}
 \color{black}
\begin{equation}
\label{eq:dwd}
    {}^{2} D^{i,l}_{k,j} = 2 \, {}^{1} D^{i}_{k} \wedge {}^{1} D^{l}_{j} - ^{2}\Delta^{i,j}_{k,l}
\end{equation}
in which the Grassmann wedge product~\cite{Mazziotti1998} is defined as
\begin{equation}
    {}^{1} D^{i}_{k} \wedge {}^{1} D^{l}_{j} = \frac{1}{2} \left ( {}^{1} D^{i}_{k} \, {}^{1} D^{l}_{j} - {}^{1} D^{i}_{j} \, {}^{1} D^{l}_{k} \right ) ,
\end{equation}
\color{black}
we obtain the particle-hole RDM ($^{2}G^{i,j}_{k,l}$) as a function of the cumulant, $^{2}\Delta^{i,l}_{k,j}$:
 \color{black}
\begin{equation}
 {}^{2}G^{i,j}_{k,l} = {}^{1} D^{i}_{j} {}^{1} D^{k}_{l} + {}^{1} D^{i}_{k} \left ( {}^{1} I^{j}_{l} - {}^{1} D^{j}_{l} \right )
    +{}^{2}\Delta^{i,j}_{k,l}
\label{eq:G2Cumulant}
\end{equation}
where the cumulant vanishes if and only if the wavefunction $| \Psi \rangle$ is a single Slater determinant.  We note that the definition of the cumulant 2-RDM in Eq.~(\ref{eq:dwd}) uses an unconventional index ordering and sign to match the index ordering and sign of the particle-hole RDM in Eq.~(\ref{eq:G2Cumulant}) rather than those of the 2-RDM~\cite{Mazziotti1998a}.  This selection of index order and sign, which facilitates the computation of the large eigenvalue associated with exciton condensation, is employed throughout the paper.
\color{black}

One \tc{trivial} large eigenvalue is always present in the particle-hole matrix, independent of exciton condensation, representing ground-state-to-ground-state projection. Therefore, the particle-hole matrix is replaced with a modified particle-hole matrix. The modified particle-hole matrix, $^{2}\tilde{G}$, removes the \tc{trivial} large eigenvalue such that a large eigenvalue present in the matrix corresponds to exciton condensation. $^{2}\tilde{G}$ is defined:
\begin{equation}
    {}^{2}\tilde{G}^{i,j}_{k,l}={}^{2}G^{i,j}_{k,l}-{}^{1}D^i_j \, {}^{1}D^l_k.
\end{equation}
For $^{2}\tilde{G}$, Garrod and Rosina \cite{Garrod1969} derived an upper bound for the eigenvalue, $\lambda_{\tilde{G}}$:
\begin{equation}
    \lambda_{\tilde{G}}\le \frac{N(r - N)}{r}.
    \label{eq:Gupperlimit}
\end{equation}
In the thermodynamic limit as $r \rightarrow \infty$, the upper bound tends to $N$.

Rearranging Eq.~(\ref{eq:G2Cumulant}) and substituting the modified particle-hole matrix for the particle-hole matrix, we obtain the cumulant RDM
\begin{equation}
\label{eq:delta}
    {}^{2}\Delta^{i,j}_{k,l} =  {}^{2}\tilde{G}^{i,j}_{k,l} - {}^{1} D^{i}_{k} \left ( {}^{1} I^{j}_{l} - {}^{1} D^{j}_{l} \right ) .
\end{equation}
\tc{The cumulant RDM differs from the modified RDM in its unconnected terms to establish the cumulant's size extensivity---a difference that is reflected in changes in its eigenvalues and eigenvectors.  For example, unless they vanish, these unconnected terms prevent the large eigenvalue of the modified particle-hole RDM from being additive for two or more exciton condensates separated by an infinite distance.}

In the case of wavefunctions exhibiting maximum long-range order, the 1-RDM becomes the identity matrix multiplied by a constant factor of $N/r$ for $N$ electrons and rank $r$ of the orbital basis set.  With this additional substitution, the cumulant RDM expression in Eq.~(\ref{eq:delta}) becomes
\begin{equation}
    \label{eq:delta2}
    {}^{2}\Delta^{i,j}_{k,l} = {}^{2}\tilde{G}^{i,j}_{k,l} - \frac{N}{r} \left ( 1 - \frac{N}{r} \right ) {}^{1} I^{i}_{k} {}^{1} I^{j}_{l}  .
\end{equation}
Given that $\lambda_{\tilde{G}} = \overrightarrow{v}^{*}\ {}^{2}\tilde{G}\ \overrightarrow{v}$, where $\lambda_{\tilde{G}}$ is the large eigenvalue of the modified particle-hole matrix and $\overrightarrow{v}$ is its eigenvector, the cumulant eigenvalue, $\lambda_\Delta$, in terms of $\lambda_{\tilde{G}}$ is found in the limit of maximum condensation by taking the expectation value of both sides of Eq.~(\ref{eq:delta2}) with respect to the eigenvector $\overrightarrow{v}$.  \tc{Note that in the limit of maximum condensation, because the cumulant matrix differs from the modified particle-hole matrix by only a scaled identity matrix in Eq.~(\ref{eq:delta2}), both the cumulant and modified particle-hole matrices share the same set of eigenvectors including $\overrightarrow{v}$.}  This yields the following expressions for $\lambda_\Delta$ in terms of $\lambda_{\tilde{G}}$:
\begin{align}
    \lambda_\Delta&=\sum_{i,j,k,l}{\overrightarrow{v}^{i}_{j}}{}^{*}\left({}^{2}\Delta^{i,j}_{k,l}\right){\overrightarrow{v}^{k}_{l}}\\
    &=\lambda_{\tilde{G}}- \frac{N}{r} \left ( 1 - \frac{N}{r} \right ) {\overrightarrow{v}^{i}_{j}}{}^{*}
    \left ( {}^{1} I^{i}_{k} {}^{1} I^{j}_{l} \right ) {\overrightarrow{v}^{k}_{l}} \\
    &= \lambda_{\tilde{G}} - \frac{N}{r} \left ( 1 - \frac{N}{r} \right ) \\
    &=\lambda_{\tilde{G}}-\frac{N(r-N)}{r^2} .
    \label{eq:cumulanteig}
\end{align}
Thus, in the thermodynamic limit as $r$ goes to infinity, $\lambda_\Delta = \lambda_{\tilde{G}}$, meaning that in the infinite basis set limit, the value of $\lambda_\Delta$ \tc{also yields the same maximum possible extent of condensation.}  From Eqs.~(\ref{eq:cumulanteig}) and (\ref{eq:Gupperlimit}), we obtain an upper bound for the value of $\lambda_\Delta$:
\begin{equation}
    \lambda_\Delta\le \frac{N(r-N)(r-1)}{r^2}
    \label{eq:cumulantlimit}
\end{equation}
for the largest eigenvalue, $\lambda_\Delta$, of the cumulant.


\section{Results}
We use two systems to examine the relationship of the large eigenvalue of the modified particle-hole matrix to that of the particle-hole cumulant. The first uses the  Lipkin model\cite{Lipkin_model, MESHKOV1965199, GLICK1965211, Perez_1988, Stein_2000, David2004, Castanos_2006, Co_2018}\color{black}, a well-known model Hamiltonian, \color{black} to \color{black} illustrate \color{black} application of the theory to an exactly solvable quantum system \color{black} that is capable of demonstrating the maximal degree of exciton condensation, which allows us to directly probe the behavior of the eigenvalue of the cumulant in the limit of maximum particle-hole condensation.  Our prior investigation into exciton condensation \cite{sager_53} has established a protocol for the construction of an exciton condensate with the maximal degree of condensation---as evinced by the largest possible eigenvalue of the modified particle-hole RDM---for a given number of fermions through use of quantum state preparation.  Here, we expand this protocol to additionally include the determination of the largest eigenvalue of the cumulant matrix.   While we use quantum simulation---i.e., classical computations that use a probabilistic approach to model the behavior of an ideal quantum computer---to compute the particle-hole and cumulant matrices of the Lipkin model for convenience as this protocol was easily adapted from prior work, these calculations can also be performed directly without the simulator in a manner similar to that described in Ref. \onlinecite{sager_coex}.  Further, having a quantum state preparation approach could allow for direct implementation on a real-world quantum simulator in order to observe how the large eigenvalue of the cumulant could change with different degrees and types of errors on real-world quantum devices.  To prove consistency with the derivation, however, we restrict our analysis to the ideal QASM simulator.  \color{black}

Second, we computationally explore a molecular system recently shown to exhibit exciton condensation using the modified particle-hole matrix. Prior results reveal that for a system of stacked benzene molecules, the degree of exciton condensation increases with the number of layers in the benzene stack, measured using the eigenvalue of the particle-hole matrix~\cite{Schouten2021a}. We use the same system to examine the eigenvalue of the particle-hole cumulant as the number of layers in the stack increased, and compare the results with those of the modified particle-hole matrix.

\subsection{The Lipkin Model}
The Lipkin model describes a quantum system composed of two, $N$-degenerate energy levels with energies of the lower and upper levels being $-\frac{\epsilon}{2}$ and $\frac{\epsilon}{2}$, respectively.  Each orbital in the lower level is particle-hole paired with an orbital in the upper level, which allows for excitations or de-excitations within a given pair but prohibits transitions between particle-hole pairs.  A Hamiltonian representing such a quantum system is given by \cite{Lipkin_model,MESHKOV1965199,GLICK1965211,Perez_1988,Stein_2000,David2004,Castanos_2006,Co_2018}
\begin{gather}
        \mathcal{H}=\frac{\epsilon}{2}\sum\limits_{\sigma=\pm1}\sigma\sum\limits_{p=1}^N\hat{a}^\dagger_{\sigma,p}\hat{a}_{\sigma,p}
        \\+\frac{\lambda}{2}\sum\limits_{\sigma=\pm1}\sum\limits_{p,q=1}^N \hat{a}^\dagger_{+\sigma,p}\hat{a}^\dagger_{+\sigma,q}\hat{a}_{-\sigma,q}\hat{a}_{-\sigma,p}
        \\+\frac{\gamma}{2}\sum\limits_{\sigma=\pm 1}\sum\limits_{p,q=1}^N\hat{a}^\dagger_{+\sigma,p}\hat{a}^\dagger_{-\sigma,q}\hat{a}_{+\sigma,q}\hat{a}_{-\sigma,p}
    \end{gather}
where $\sigma$ and $p$ are quantum numbers representing the level ($\sigma = \pm 1$) and the index of a pair and where $\gamma$ and $\lambda$ tune the strengths of the single and double excitations and de-excitations.

As has been shown in Ref.~\cite{sager_53}, an exciton condensate can be prepared on a quantum device by entangling a state representing all $N$ particles occupying the lower level with a state in which all $N$ particles occupy the upper level.  Treating each qubit as a one-particle orbital, such a preparation can be achieved as shown in the Appendix and yields wavefunctions of the form
\color{black}
\begin{equation}
    |\Psi\rangle =  \frac{1}{\sqrt{2}}  \left ( |1\dots10\dots0\rangle+|0\dots01\dots 1\rangle \right )
    \label{eq:wavefunction}
\end{equation}
\color{black}
where the first $N$ qubits correspond to the lower level and the last $N$ qubits correspond to the upper level.

For this case, the upper bound is clearly defined since $r = 2N$. Consequently, $\lambda_\Delta$ and $\lambda_{\tilde{G}}$ are related by:
\begin{equation}
   \lambda_\Delta=\lambda_{\tilde{G}}-\frac{1}{4}.
   \label{eq:specificrelationship}
\end{equation}
The theoretical maximum value of $\lambda_{\tilde{G}}$ is then $N/2$ so the upper bound of $\lambda_\Delta$ is:
\begin{equation}
    \lambda_\Delta\le \frac{N}{2}-\frac{1}{4}.
\label{eq:specificlimit}
\end{equation}

The largest eigenvalue of the modified particle-hole matrix ($\lambda_{\tilde{G}}$) and the largest eigenvalue of the cumulant ($\lambda_\Delta$) corresponding to these states for various particle numbers ($N$) are shown in Fig.~\ref{fig:lipkin}.  (See the Appendix for details on how these values were obtained via post-measurement tomography.)  As this figure demonstrates, the signature of exciton condensation is indeed large---$\lambda_{\tilde{G}}>1$---and increases linearly with system size---consistent with the theoretical maximum of $\frac{N}{2}$.  Moreover, the large eigenvalue of the cumulant---$\lambda_\Delta$---is additionally large, demonstrating that this eigenvalue can be utilized as a measure of condensation.  The large eigenvalue in the cumulant demonstrates that the condensation and its long-range order arise from a non-trivial entanglement that increases linearly with the system size. Further, in this system demonstrating maximal exciton condensation, the eigenvalue of the cumulant exactly matches that from the derivation shown in the Theory section.

\begin{figure}
    \centering
    \includegraphics[width=8cm]{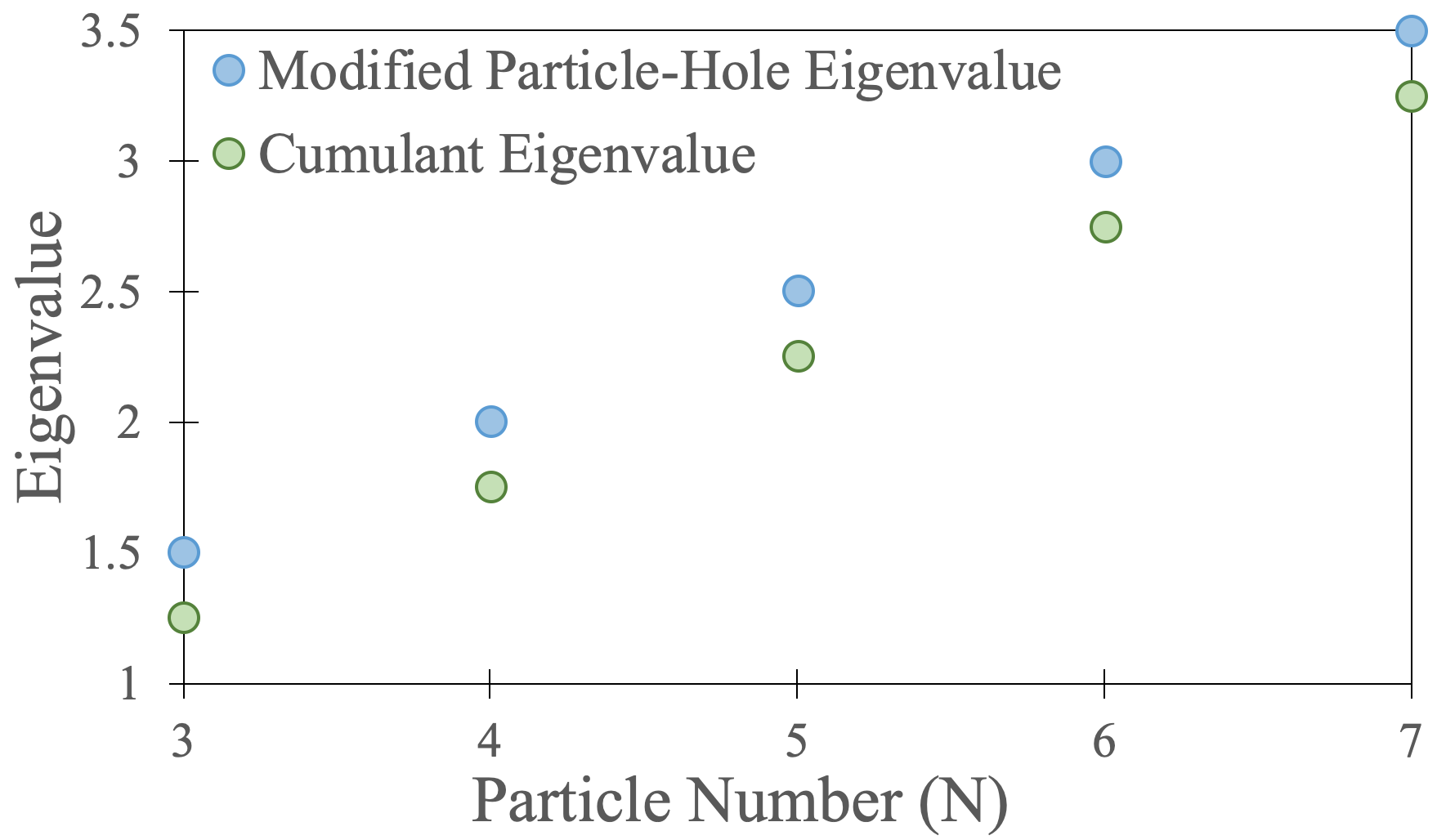}
    \caption{\label{fig:lipkin} A figure detailing the \color{black} signature of the extent of exciton condensation---i.e., the largest eigenvalue of the modified particle-hole matrix  \color{black}($ \lambda_{\tilde{G}}$)---as well as the proposed size-extensive signature of exciton condensation---i.e., the largest eigenvalue of the cumulant ($\lambda_\Delta$) matrix---for implementation of the wavefunction detailed in Eq.~(\ref{eq:wavefunction}) using IBM's QASM simulator for a variety of particle numbers ($N$)  is shown. Note that as this wavefunction has maximal character of exciton condensation, we expect both signatures to be maximal ($\lambda_{\Tilde{G}}=\frac{N}{2}, \ \lambda_\Delta=\lambda_{\Tilde{G}}-\frac{1}{4}=\frac{N}{2}-\frac{1}{4}$), which is consistent with the data obtained from simulation, verifying our predicted signature of condensation. \color{black}}
\end{figure}

\subsection{Benzene Stacks}

\begin{figure}[tbh!]
    \centering
    \includegraphics[width=8cm]{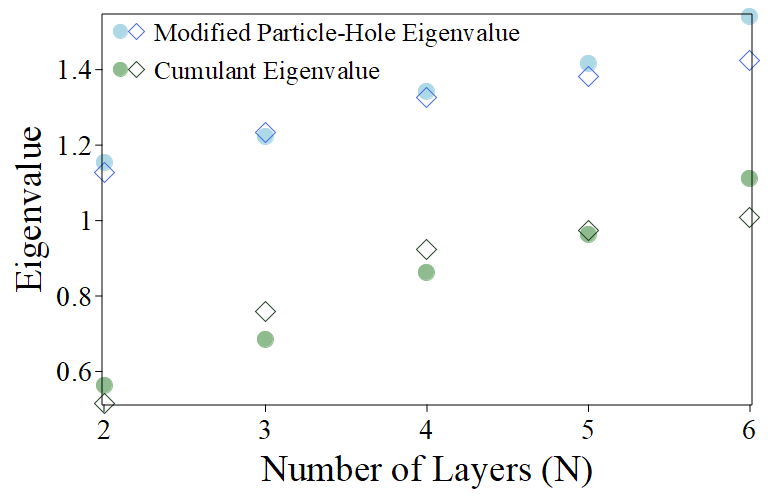}
    \caption{Comparison of the calculated eigenvalues of the modified particle-hole and cumulant matrices for benzene stacks. Circles represent the STO-6G basis set, and diamonds are the cc-pVDZ basis set.}
    \label{fig:benzene}
\end{figure}

To examine the cumulant measure of exciton condensation in a molecular system, we \textcolor{black}{use variational 2-RDM theory \cite{Mazziotti1998, Mazziotti2012, Mazziotti2004, mazziotti_2016, nakata_nakatsuji_ehara_fukuda_nakata_fujisawa_2001, M2007} to computationally probe} stacks of benzene molecules ranging from two to six layers \textcolor{black}{(see Appendix for calculation details)}. Each layer was separated by 2.5~\AA, consistent with previous \textcolor{black}{computational} studies \cite{Safaei2018}. Previous results \cite{Schouten2021a} obtained from the eigenvalues of the modified particle-hole matrix show that exciton condensation occurs in all stacks---indicated by an eigenvalue greater than one---and the extent increases with the number of layers in the stack. As seen in Figure \ref{fig:benzene} and Table \ref{tab:benzeneeig}, the results for the cumulant follow the same monotonically increasing trend. \textcolor{black}{Results are given for calculations with two different sets of basis functions: STO-6G, \cite{Hehre1969} a minimal Slater-type orbital basis set, and cc-pVDZ, \cite{Dunning1989} a double-zeta correlation consistent basis set. For both basis sets the eigenvalues follow the same increasing trend, indicating the trend is not basis set dependent.} Consistent with the theory presented above, the eigenvalues of the cumulant are all less than those of the modified particle-hole matrix. The benzene system does not reach the theoretical maximum, so there is greater difference between $\lambda_{\tilde{G}}$ and $\lambda_\Delta$ than for the Lipkin model, making several of the values of $\lambda_\Delta$ less than one. Despite the fact that the eigenvalues for the cumulant are less than one, the increase in cumulant eigenvalues following the pattern of the modified particle-hole matrix suggests that both matrices and their eigenvalues recover the same correlation.  Moreover, as in the Lipkin model, the large eigenvalue of the cumulant matrix indicates that the condensation is not a consequence of a mean-field long-range order but rather a non-diagonal long-range order arising from non-trivial entanglement of the excitons. \tc{Unlike the Lipkin model, we do not expect to saturate the bound. Saturation of the bound would correspond to participation of all particles in condensation, which we do not expect in benzene.}

\begin{table}[tbh!]

    \caption{Eigenvalues of the modified particle-hole and cumulant matrices for benzene stacks.}
    \label{tab:benzeneeig}
     \centering
    \begin{ruledtabular}
    \begin{tabular}{c c c c c}
                & \multicolumn{2}{c}{cc-pVDZ} & \multicolumn{2}{c}{STO-6G}\\ \cline{2-3} \cline{4-5}
      Layers   & $\lambda_{\tilde{G}}$ & $\lambda_\Delta$ & $\lambda_{\tilde{G}}$ & $\lambda_\Delta$ \\
      \hline
      2   &  1.1273 & 0.5166 & 1.1542 & 0.5618 \\
      3 & 1.2338 & 0.7598 & 1.2232 & 0.6843 \\
      4 & 1.3275 & 0.9240 & 1.3418 & 0.8635 \\
      5 & 1.3812 & 0.9735 & 1.4175 & 0.9625 \\
      6 & 1.4254 & 1.0074 & 1.5422 & 1.1128 \\
    \end{tabular}
    \end{ruledtabular}

\end{table}

To compare the results of the two matrices, we visualize the ``exciton density" for both the cumulant and the modified particle-hole matrices. The visualization constructs a picture of the probabilistic density of the exciton by using the eigenvectors of the \tc{associated} matrix to calculate the hole probability for a given exciton as a function of a particle localized in a specific atomic orbital. Details of the calculation are found in the Appendix. Figure \ref{fig:orbitals} shows the densities for each matrix for the four-layer stack, although the same conclusions \textcolor{black}{apply} to the other stacks as well. Both densities show similar delocalization of the hole over the layers, with the only differences being slight variations in the size of the densities. Since both systems show essentially the same density, it is reasonable to conclude that the cumulant recovers the same electron-hole correlation as the modified particle-hole matrix.

\begin{figure}[tbh!]
    \centering
    \includegraphics[width=8cm]{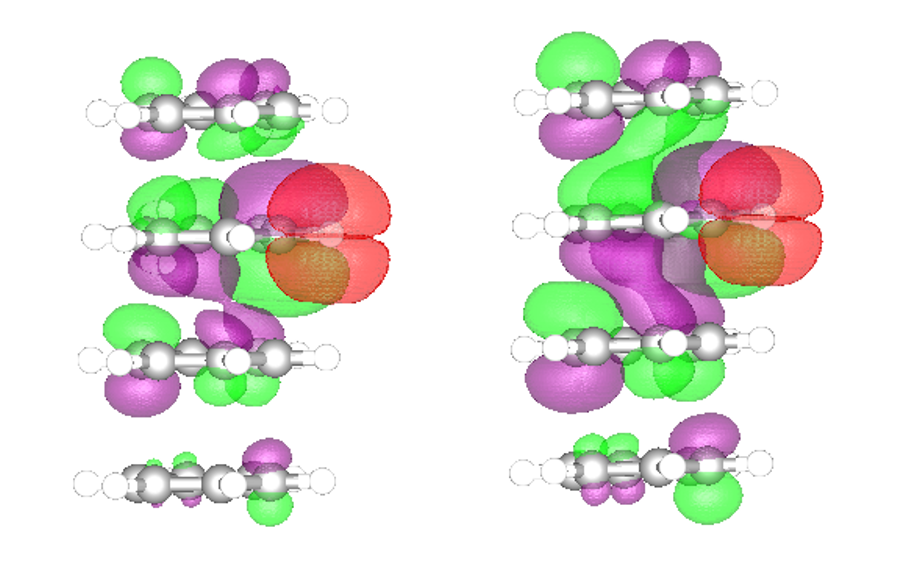}
    \caption{Comparison of the ``exciton densities" calculated using the modified particle-hole (left) and cumulant (right) matrices.}
    \label{fig:orbitals}
\end{figure}

\section{Discussion and Conclusions}

Exciton condensation has a signature in the particle-hole RDM with a related signature in the particle-hole cumulant. We derived the relationship between the eigenvalue of the particle-hole RDM and that of the particle-hole cumulant, showing an upper bound exists for the value of the cumulant eigenvalue. We found a general relationship between the eigenvalues of the particle-hole RDM and cumulant such that the largest eigenvalue of the particle-hole RDM is greater than or equal to that of the particle-hole cumulant with equality in the thermodynamic limit.  The large eigenvalue in the cumulant matrix reveals the role of entanglement in the off-diagonal long-range order of the condensation.

We explored the theoretical result using the Lipkin model to prepare an exciton condensate on a quantum simulator. The results from the Lipkin model are consistent with the theoretical maximum predicted by the theory, showing a linear increase in the eigenvalue with the number $N$ of particles corresponding to eigenvalues of $N/2$ for the particle-hole RDM. Eigenvalues of the cumulant are large, but less than those of the particle-hole RDM by $1/4$, consistent with the theoretical prediction. Additionally, we showed for a system of stacked benzene molecules that the eigenvalues of both the particle-hole RDM and the cumulant increase with the number of layers. Furthermore, the modeled ``exciton densities” for the particle-hole RDM and cumulant of the stacked benzene molecules are structurally similar. Although the eigenvalues of the cumulant are not large, the growth in eigenvalues corresponding to that of the particle-hole RDM and the similarity of the densities indicate the same correlation corresponding to exciton condensation in both matrices.

The cumulant is calculable directly from the 2-RDM. While the cumulant obtained directly from the 2-RDM is associated with particle-particle correlations, particle-hole correlations are contained in the same cumulant and obtained simply by switching the indices. Thus, the cumulant contains the information associated with both particle-hole and particle-particle correlations. As previous work revealed a measure of off-diagonal long-range order in the cumulant for the particle-particle case, from this work we conclude that the 2-RDM cumulant offers a metric for off-diagonal long-range order associated with condensation of both particle-hole and  fermion-pairs. Significantly, \textcolor{black}{a recent theoretical study revealed the possibility for the existence of} dual condensates of excitons and  fermion-pairs, where both types of condensation are present in a single condensate phase in the material \cite{Sager2020}.  \color{black} Such a dual condensate has subsequently been prepared and probed on an experimental quantum device \cite{Sager_QCDUAL}. \color{black} The ability of the cumulant to serve as a metric for both condensation phenomena has great potential for applications involving dual condensates.


The potential for application of exciton condensation to efficient energy transport seems evident; however, the phenomenon proves difficult to realize experimentally. Accessible computational measures of exciton condensation could consequently prove critical to eventually realizing this potential. Theoretical measures, like those of the particle-hole RDM and cumulant, provide the ability to explore exciton condensation in a wide array of systems for potentially practical realization. The nature of the cumulant is particularly well-suited to this task, as the cumulant provides a metric for exciton condensation that is size-extensive in both the finite and the thermodynamic limits. Consequently, in the case of two condensates at infinite separation, the cumulant would appropriately measure both condensates, whereas in the particle-hole RDM the separation of the two may become obscured. \tc{Moreover, the size extensive properties of the cumulant measure may be especially important in any process that changes the size of the condensate, i.e. a process that separates or combines condensates. The eigenvalue of two separate condensates is additive in the cumulant matrix, but not in the particle-hole RDM.} Because the cumulant scales linearly with system size, it may also better capture the off-diagonal long-range order associated with exciton condensation, making it a more natural metric than the particle-hole RDM. The cumulant signature of exciton condensation provides a valuable tool for computational explorations towards uncovering systems and materials capable of exhibiting exciton condensation.

\begin{acknowledgments}
D.A.M. gratefully acknowledges support from the De-
partment of Energy, Office of Basic Energy Sciences under
Grant No. DE-SC0019215, the ACS Petroleum Research
Fund under Grant No. PRF No. 61644-ND6, and the U.S.
National Science Foundation under Grant No. CHE-1565638
and Grant No. CHE-2035876. L.M.S.-S. also acknowledges
support from the U. S. National Science Foundation under
Grant No. DGE-1746045.
\end{acknowledgments}

\appendix
\section*{Appendix}
\addcontentsline{toc}{section}{Appendix}
\renewcommand{\thesubsection}{Appendix \Alph{subsection}}
We include details on the quantum algorithms used to prepare the qubit states presented in the article; the quantum tomography of the modified  particle-hole reduced density matrix; the quantum tomography of the portion of the cumulant corresponding to the large eigenvalue, relevant details on the simulated quantum backend employed and details of methods used for molecular calculations.

\subsection{State Preparation}
\begin{figure}[tbh!]
    \includegraphics[width=8cm]{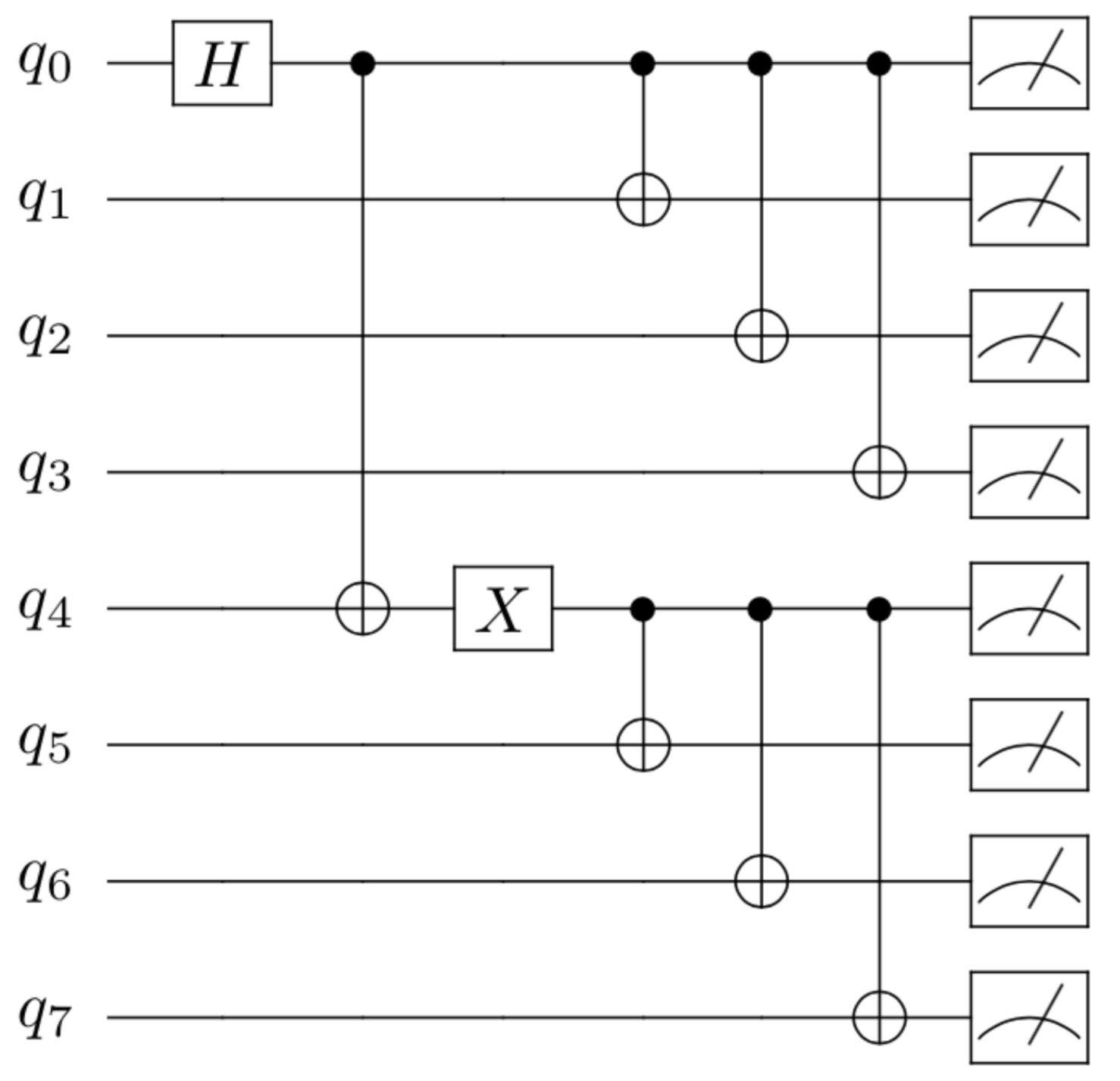}
    \caption{\label{fig:N4circ} A schematic demonstrating the quantum state preparation described in Eq. (\ref{eq:stateprep}) for $N=4,r=8$ which yields Eq. (\ref{eq:wavefunction}).\color{black}. Note that $q_i$ corresponds to a qubit with index $i$, $H$ corresponds to the implementation of the Hadamard gate, $X$ corresponds to the implementation of the $X$ gate, and each $\bullet$ with a related $\bigoplus$ corresponds to a CNOT gate with the control qubit given by the $\bullet$ and the target qubit given by the $\bigoplus$. \color{black}}
\end{figure}

Our previous work \cite{sager_53} has demonstrated that maximal exciton condensate character (as evinced by $\lambda_{\tilde{G}}=\frac{N}{2}$) can be obtained by
\begin{equation}
    |\Psi\rangle=\left(\prod\limits_{i=1}^{N-1}C_N^{N+i}\right)\left(\prod\limits_{i=1}^{N-1}C_0^{i}\right)X_NC_0^NH_0|\Psi_0\rangle
\label{eq:stateprep}
\end{equation}
when each individual qubit is treated as a one-particle orbital and where $H_i$ corresponds to a Hadamard gate on qubit $i$, $C_i^j$ corresponds to a CNOT gate with qubits $i$ and $j$ representing the target and control qubits,  with $X_i$ corresponding to a $X$ gate applied to the $i^{th}$ qubit, \color{black} and with $|\Psi_0\rangle$ corresponding to the conventional initial state on a quantum device in which all qubits are in the $|0\rangle$ state. \color{black}  Such a preparation yields a preparation given by Eq. (\ref{eq:wavefunction}) and is represented pictorially in Fig. \ref{fig:N4circ} for a system composed of $N=4$ particles in $r=8$ orbitals.

\subsection{Quantum Tomography}
\begin{widetext}
\subsubsection{The Modified Particle-Hole Reduced Density Matrix}
The signature of the particle-hole reduced density matrix $\lambda_{\tilde{G}}$ associated with exciton condensation is computed analogously to the methodology described in Ref.~\cite{Dual2022}.  In summary, the overall particle-hole matrix associated with the large eigenvalue is given by the following tiling
\begin{equation*}
\begin{array}{|c|c|c|c|}
	\hline
	{\scriptstyle  p=0,q=0} & {\scriptstyle  p=0, q=1} & \cdots & {\scriptstyle  p=0,q=\frac{N}{2}-1} \\\hline
	{ \scriptstyle p=1,q=0} & { \scriptstyle p=1, q=1} &  \cdots & { \scriptstyle p=1,q=\frac{N}{2}-1} \\\hline
	\vdots & \vdots & \ddots & \vdots \\\hline
	{ \scriptstyle p=\frac{N}{2}-1,q=0 }& { \scriptstyle p=\frac{N}{2}-1, q=1} & \cdots &{ \scriptstyle  p=\frac{N}{2}-1,q=\frac{N}{2}-1} \\\hline
	\end{array}
\end{equation*}
with each tile being represented by
\begin{equation*}
\begin{array}{c|cccc}
     &  { \hat{a}_{q}^\dagger\hat{a}_{q}} &  { \hat{a}_{q+4}^\dagger\hat{a}_{q}} &  { \hat{a}_{q}^\dagger\hat{a}_{q+4}} &  { \hat{a}_{q+4}^\dagger\hat{a}_{q+4}}\\\hline
   { \hat{a}_{p}^\dagger\hat{a}_{p}} & {  \hat{a}_{p}^\dagger\hat{a}_{p}\hat{a}_{q}^\dagger\hat{a}_{q}}  &  { \hat{a}_{p}^\dagger\hat{a}_{p}\hat{a}_{q+4}^\dagger\hat{a}_{q}} &  { \hat{a}_{p}^\dagger\hat{a}_{p}\hat{a}_{q}^\dagger\hat{a}_{q+4}} &  { \hat{a}_{p}^\dagger\hat{a}_{p}\hat{a}_{q+4}^\dagger\hat{a}_{q+4}}\\
  {  \hat{a}_{p}^\dagger\hat{a}_{p+4}} &  { \hat{a}_{p}^\dagger\hat{a}_{p+4}\hat{a}_{q}^\dagger\hat{a}_{q}}  &  { \hat{a}_{p}^\dagger\hat{a}_{p+4}\hat{a}_{q+4}^\dagger\hat{a}_{q}} &  { \hat{a}_{p}^\dagger\hat{a}_{p+4}\hat{a}_{q}^\dagger\hat{a}_{q+4}} &  { \hat{a}_{p}^\dagger\hat{a}_{p+4}\hat{a}_{q+4}^\dagger\hat{a}_{q+4}}\\
   { \hat{a}_{p+4}^\dagger\hat{a}_{p}} &  { \hat{a}_{p+4}^\dagger\hat{a}_{p}\hat{a}_{q}^\dagger\hat{a}_{q}}  &  { \hat{a}_{p+4}^\dagger\hat{a}_{p}\hat{a}_{q+4}^\dagger\hat{a}_{q}} &  { \hat{a}_{p+4}^\dagger\hat{a}_{p}\hat{a}_{q}^\dagger\hat{a}_{q+4} }& {  \hat{a}_{p+4}^\dagger\hat{a}_{p}\hat{a}_{q+4}^\dagger\hat{a}_{q+4}}\\
   { \hat{a}_{p+4}^\dagger\hat{a}_{p+4}} &  { \hat{a}_{p+4}^\dagger\hat{a}_{p+4}\hat{a}_{q}^\dagger\hat{a}_{q}}   &  { \hat{a}_{p+4}^\dagger\hat{a}_{p+4}\hat{a}_{q+4}^\dagger\hat{a}_{q}} &  { \hat{a}_{p+4}^\dagger\hat{a}_{p+4}\hat{a}_{q}^\dagger\hat{a}_{q+4}} &  { \hat{a}_{p+4}^\dagger\hat{a}_{p+4}\hat{a}_{q+4}^\dagger\hat{a}_{q+4}}.
\end{array}
\label{eq:G2details}
\end{equation*}
The extraneous large eigenvalue of the particle-hole matrix corresponding to a ground-state-to-ground state transition is then removed by subtracting off
\begin{equation*}
\begin{array}{c|cccc}
     &  {\hat{a}_{q}^\dagger\hat{a}_{q}} &  {\hat{a}_{q+4}^\dagger\hat{a}_{q}} &  {\hat{a}_{q}^\dagger\hat{a}_{q+4}} &  {\hat{a}_{p+4}^\dagger\hat{a}_{p+4}}\\\hline
   { \hat{a}_{p}^\dagger\hat{a}_{p}} &  { {}^{1}D_p[0,0]{}^{1}D_q[0,0]} &   { {}^{1}D_p[0,0]{}^{1}D_q[0,1]} &   { {}^{1}D_p[0,0]{}^{1}D_q[1,0]} &  { {}^{1}D_p[0,0]{}^{1}D_q[1,1]} \\
  {  \hat{a}_{p}^\dagger\hat{a}_{p+4}} &  { {}^{1}D_p[0,1]{}^{1}D_q[0,0]} &   { {}^{1}D_p[0,1]{}^{1}D_q[0,1]} &   { {}^{1}D_p[0,1]{}^{1}D_q[1,0]} &  { {}^{1}D_p[0,1]{}^{1}D_q[1,1]}  \\
   { \hat{a}_{p+4}^\dagger\hat{a}_{p}} &  { {}^{1}D_p[1,0]{}^{1}D_q[0,0]} &   { {}^{1}D_p[1,0]{}^{1}D_q[0,1]} &   { {}^{1}D_p[1,0]{}^{1}D_q[1,0]} &  { {}^{1}D_p[1,0]{}^{1}D_q[1,1]} \\
   { \hat{a}_{p+4}^\dagger\hat{a}_{p+4}} &  { {}^{1}D_p[1,1]{}^{1}D_q[0,0]} &   { {}^{1}D_p[1,1]{}^{1}D_q[0,1]} &   { {}^{1}D_p[1,1]{}^{1}D_q[1,0]} &  { {}^{1}D_p[1,1]{}^{1}D_q[1,1]}
\end{array}
\end{equation*}
from each tile where ${}^{1}D_p$ corresponds to a one-density matrix given by
\begin{equation*}
{}^{1}D_p=\begin{array}{c|cc}
	& {\hat{a}_{p}}& {\hat{a}_{p+4}} \\\hline
	{\hat{a}^\dagger_{p}} & {\hat{a}^\dagger_{p}\hat{a}_{p}} & {\hat{a}^\dagger_{p}\hat{a}_{p+4}} \\
	{\hat{a}^\dagger_{p+4}} & {\hat{a}^\dagger_{p+4}\hat{a}_{p}} & {\hat{a}^\dagger_{p+4}\hat{a}_{p+4}}
	\end{array}
\end{equation*}

This matrix is computed directly from the ``counts'' data obtained directly from the quantum device---i.e., how many times each basis state was obtained upon probing the state preparation---by representing the fermionic creation and annihilation operators as the corresponding qubit operators and by evaluating the expectation tensor product of the appropriate creation and annihilation operators for each element of the matrix.

The signature of excitation condensation ($\lambda_{\tilde{G}}$) can then be directly computed by solving the following eigenvalue equation
\begin{equation}
    {}^{2}\Tilde{G}\overrightarrow{v}_{\tilde{G}}^{\ i}=\epsilon_{\tilde{G}}^i\overrightarrow{v}_{\tilde{G}}^{\ i}
\end{equation}
with the signature corresponding the largest eigenvalue (the maximum $\epsilon_{\tilde{G}}^i$).

\subsubsection{The Cumulant}
As the cumulant can be directly computed from the unmodified particle-hole matrix and the one-particle density matrices described in the previous section, each element of the cumulant corresponding to the elements of the modified particle-hole density matrix can be computed directly from Eq.~(\ref{eq:delta}).  Thus, the largest eigenvalue of the cumulant ($\lambda_\Delta$) can be obtained from solving the eigenvalue equation
\begin{equation}
    {}^{2}\Delta \overrightarrow{v}_\Delta^{\ i}=\epsilon_\Delta^i\overrightarrow{v}_\Delta^{\ i}
\end{equation}
and obtaining the maximum $\epsilon_\Delta^i$.

\end{widetext}

\subsection{Simulated Quantum Device Specifications}
Throughout this work, we have employed Qiskit Aer's qasm simulator from the open source Qiskit software stack, which is available online.  See the Qiskit documentation \cite{Qiskit,qiskit_2021} for detailed specifications of this quantum device simulator.

\subsection{Computational Methods}
The 2-RDM for the benzene stacks is calculated directly from the molecular structure using a variational method \cite{Mazziotti1998, Mazziotti2012, Mazziotti2004, mazziotti_2016, nakata_nakatsuji_ehara_fukuda_nakata_fujisawa_2001, M2007}. N-representability conditions constrain the 2-RDM which require the 2-RDM, two-hole RDM, and particle-hole RDM to be positive semi-definite. The STO-6G and cc-pVDZ basis sets are used for benzene stack calculations, and active spaces are determined based on the number of $\pi$ electrons: [12,12] for the two-layer, [18,18] for the three-layer, etc.

The 2-RDM can be used to obtain the particle-hole RDM by the linear mapping in Eq.~(\ref{eq:g}).  From the particle-hole RDM, the modified particle-hole RDM and the particle-hole cumulant can be obtained as described in the Theory section. Eigenvalues ($\lambda_i$) and eigenvectors ($\overrightarrow{v}_i$) of these matrices are calculated using an eigenvalue optimization:
\begin{equation}
    M\overrightarrow{v}_i = \lambda_i\overrightarrow{v}_i
\end{equation}
where $M$ can be either the modified particle-hole matrix or the particle-hole cumulant.

The visualization of the ``exciton density" plots the hole probability as a function of a specific particle location using a matrix of molecular orbitals in terms of atomic orbitals, $M_\mathrm{{MO,AO}}$, calculated:
\begin{equation}
    M_\mathrm{{AO,MO}} = (M^{T}_\mathrm{{MO,AO}})^{-1}.
\end{equation}
Active orbitals are isolated from the main matrix as a submatrix, and the eigenvector of the large eigenvalue is reshaped as a matrix in the basis of the submatrix. The eigenvector matrix, $V_\mathrm{{max}}$, is used to create a matrix representation of electron atomic orbitals in terms of the hole coefficients corresponding to contributions to other molecular orbitals:
\begin{equation}
    (M^\mathrm{{active}}_\mathrm{{AO,MO}})(V_\mathrm{{max}})(M^\mathrm{{active}}_\mathrm{{AO,MO}})^{T}.
\end{equation}

	\bibliography{benzene_exciton}

\end{document}